\documentclass[a4paper,11pt]{article}
\usepackage[DIV12]{typearea}
\usepackage{longtable}
\usepackage{booktabs}
\usepackage{multirow}
\usepackage{amsmath}
\usepackage{amssymb}
\usepackage{bbm}
\usepackage[utf8]{inputenc}
\usepackage{pdflscape}
\usepackage{xspace}
\usepackage[caption=false]{subfig}
\usepackage{nicefrac}
\usepackage{graphicx}
\usepackage{cite}  
\usepackage[small]{caption2}

\usepackage[all,cmtip]{xy}
\newlength{\xywd}
\newcommand{\xyrightarrow}[2][]{%
  \sbox{0}{$\scriptstyle#1$}%
  \xywd=\wd0
  \sbox{0}{$\scriptstyle#2$}%
  \ifdim\wd0>\xywd \xywd=\wd0 \fi
  \xymatrix@C\dimexpr\xywd+1em\relax{{}\ar[r]^{#2}_{#1}&{}}%
}

\DeclareMathOperator{\re}{Re}
\DeclareMathOperator{\im}{Im}

\newcommand{\Z}[1]{\ensuremath{\mathbbm{Z}_{#1}}} 
\newcommand{\SO}[1]{\ensuremath{\mathrm{SO}(#1)}}

\newcommand{\SL}[1]{\ensuremath{\mathrm{SL}(#1)}}

\newcommand{\E}[1]{\ensuremath{\mathrm{E}_{#1}}}
\newcommand{\I}{\mathrm{i}}
\newcommand{\Id}{\mathbbm{1}}

\newcommand{\CP}{\ensuremath{\mathcal{CP}}\xspace}
\newcommand{\x}{\ensuremath{\times}}

\newcommand{\Sp}[1]{\ensuremath{\mathrm{Sp}(#1,\mathbbm{Z})}}
\newcommand{\dd}{\ensuremath{\mathrm{d}}}

\newcommand{\W}[2]{\ensuremath{\hat{W}\!\left(\!\begin{smallmatrix}#1\\ #2\end{smallmatrix}\!\right)}}
\newcommand{\M}[2]{\ensuremath{M\!\left(\!\begin{smallmatrix}#1\\ #2\end{smallmatrix}\!\right)}}
\newcommand{\s}[2]{\ensuremath{\left(\!\begin{smallmatrix}#1\\ #2\end{smallmatrix}\!\right)}}

\usepackage{xcolor}

\advance \headheight by 3.0truept       
\usepackage[pdftex]{hyperref}
\hypersetup{
    pdftitle = {Siegel modular flavor group and CP from string theory},
    pdfauthor = {Baur, Kade, Nilles, Ramos-Sanchez, Vaudrevange}
}

\begin{document}

\begin{titlepage}

\begin{flushright}
\normalsize{TUM-HEP 1307/20}
\end{flushright}

\vspace*{1.0cm}

\begin{center}
{\Large\textbf{\boldmath Siegel modular flavor group and \CP from string theory}\unboldmath}

\vspace{1cm}

\textbf{Alexander Baur$^{a,c}$, Moritz Kade$^{a}$,}
\textbf{Hans Peter Nilles$^{b}$, Sa\'ul Ramos--S\'anchez$^{c}$, Patrick K.S. Vaudrevange$^{a}$}
\\[8mm]
\textit{$^a$\small Physik Department T75, Technische Universit\"at M\"unchen,\\ James-Franck-Stra\ss e 1, 85748 Garching, Germany}
\\[2mm]
\textit{$^b$\small Bethe Center for Theoretical Physics and Physikalisches Institut der Universit\"at Bonn,\\ Nussallee 12, 53115 Bonn, Germany}
\\[2mm]
\textit{$^c$\small Instituto de F\'isica, Universidad Nacional Aut\'onoma de M\'exico,\\ POB 20-364, Cd.Mx. 01000, M\'exico}
\end{center}

\vspace{1cm}

\vspace*{1.0cm}

\begin{abstract}
We derive the potential modular symmetries of heterotic string theory. For a toroidal 
compactification with Wilson line modulus, we obtain the Siegel modular group $\mathrm{Sp}(4,\Z{})$ 
that includes the modular symmetries $\mathrm{SL}(2,\Z{})_T$ and $\mathrm{SL}(2,\Z{})_U$ (of the 
``geometric'' moduli $T$ and $U$) as well as mirror symmetry. In addition, string theory provides a 
candidate for a \CP-like symmetry that enhances the Siegel modular group to $\mathrm{GSp}(4,\Z{})$.
\end{abstract}

\end{titlepage}

\newpage

\section{Introduction}

Modular symmetries might play an important role for a description of the flavor structure in 
particle physics~\cite{Feruglio:2017spp}. In string theory, modular transformations appear as the 
exchange of winding and momentum (Kaluza-Klein) modes in compactified extra dimensions, combined 
with a nontrivial transformation of the moduli. In the application to flavor symmetries, these 
moduli play the role of flavon fields that are responsible for the spontaneous breakdown of flavor 
and \CP symmetries. While string theory requires six compact space dimensions with many moduli, the 
explicit discussion in flavor physics has, up to now, mainly concentrated on two compact extra 
dimensions and few geometric moduli (see e.g.\ ref.~\cite{deMedeirosVarzielas:2019cyj}). In the 
top-down discussion, this included i) the $\mathbbm{T}^2/\Z{3}$ orbifold with K\"ahler modulus $T$ 
(and frozen complex structure modulus $U$)~\cite{Baur:2019kwi,Baur:2019iai,Nilles:2020kgo} subject 
to the modular group $\SL{2,\Z{}}_T$ and ii) the $\mathbbm{T}^2/\Z{2}$ orbifold with $T$ and $U$ 
moduli with a corresponding modular group $\SL{2,\Z{}}_T \times \SL{2,\Z{}}_U$ combined with a 
mirror symmetry that interchanges $T$ and $U$~\cite{Baur:2020jwc}.

The present paper performs a next step towards a more exhaustive discussion of the 
``many-moduli-case''. Our results are based on the observation that string theory includes more 
moduli beyond the (geometric) $T$- and $U$-moduli in form of Wilson lines connected to gauge 
symmetries in extra dimensions. Modular transformations act nontrivially on these Wilson lines and 
require a modified geometric interpretation. In the present paper, we illustrate this situation 
for compactifications on two-tori and the corresponding transformation of the Narain lattice in 
heterotic string theory. Our main results are:
\begin{itemize}
\item Wilson line moduli lead to an enhancement of modular flavor symmetries,
\item for the case of two compactified dimensions, this leads to the appearance of the Siegel 
modular group \Sp{4}, which includes $\SL{2,\Z{}}_T \times \SL{2,\Z{}}_U$ as well as mirror 
symmetry,
\item a generalized geometric interpretation of the origin of these symmetries is given through an 
auxiliary Riemann surface of genus 2 (see figure~\ref{fig:tori}) that combines the metric and gauge 
moduli in a common setting\footnote{This interpretation was first anticipated in the discussion of 
gauge threshold corrections in heterotic string theory~\cite{Mayr:1995rx}.}, and
\item a candidate \CP-like symmetry naturally appears in string models with two compact 
dimensions; interestingly, this symmetry also arises in a bottom-up discussion as an outer 
automorphism of the Siegel modular group, extending it to $\mathrm{GSp}(4,\Z{})$.
\end{itemize}

The paper is organized as follows. In section~\ref{sec:Sp2gZ}, we introduce the Siegel modular 
group \Sp{2g}. Specific properties and subgroups are illustrated for the genus 2 case \Sp{4}, where 
the subgroups include $\SL{2,\Z{}}_T \times \SL{2,\Z{}}_U$ and mirror symmetry. Besides the $T$- 
and $U$-moduli, the Siegel modular group acts on a third modulus $Z$. In 
section~\ref{sec:OriginOfSP4Z}, we relate this third modulus to Wilson lines in heterotic string 
theory. We introduce the $2D+16$-dimensional Narain lattice and its outer automorphism 
$\mathrm{O}_{\hat\eta}(D,D+16,\Z{})$ and then specialize on $D=2$ with a nontrivial Wilson line. 
The subgroup $\mathrm{O}_{\hat\eta}(2,3,\Z{})$ of $\mathrm{O}_{\hat\eta}(2,2+16,\Z{})$ can be 
mapped to \Sp{4} as given explicitly in table~\ref{tab:Mapping}. This allows for a connection to 
the recent bottom-up approach of Ding, Feruglio and Liu~\cite{Ding:2020zxw}. Their ``third'' 
modulus can thus be realized as a Wilson line in heterotic string theory. In addition, string 
constructions admit a \CP-like symmetry, which appears at the same footing as all discrete 
(traditional and modular) symmetries. In section~\ref{sec:GSp}, we show that this \CP-like 
symmetry also appears naturally from a bottom-up perspective: It corresponds to an outer 
automorphism of the Siegel modular group extending it to the general symplectic group 
$\mathrm{GSp}(4,\Z{})$. Conclusions and outlook are given in section~\ref{sec:conclusions}. 
Finally, some technical details are discussed in two appendices.

\section{\boldmath The Siegel modular group \Sp{2g} \unboldmath}
\label{sec:Sp2gZ}

The symplectic group over the integers \Sp{2g} (also called the Siegel modular group of genus $g$) 
is the group of linear transformations $M$ which preserve a skew-symmetric bilinear form $J$, i.e.
\begin{equation}\label{eq:DefSp2gZ}
\Sp{2g} ~:=~ \left\lbrace M\in \Z{}^{2g\times 2g} ~\vert~ M^{\mathrm{T}}J M = J \right\rbrace\;.
\end{equation}
Here, $J$ is given as
\begin{equation}
  J ~:=~ \begin{pmatrix} 0& \Id_g \\ -\Id_g & 0\end{pmatrix}\;,
\label{eq:J}
\end{equation}
and $\Id_g$ is the $g$-dimensional identity matrix. 

As reviewed in appendix~\ref{app:Genusg}, there exists a natural action of \Sp{2g} on a symmetric 
$g \times g$-dimensional matrix called $\Omega \in \mathbbm H_g$, where the Siegel upper half plane 
$\mathbbm H_g$ is defined as
\begin{equation}
\mathbbm H_g ~:=~ \left\{\Omega\in \mathbbm{C}^{g\times g}~|~ \Omega^\mathrm{T}=\Omega\;,\; \im\Omega>0 \right\}\;.
\end{equation}
Hence, $\Omega$ contains $g\times(g+1)/2$ complex numbers that are called moduli. In more detail, 
one splits $M \in \Sp{2g}$ into $g \times g$-dimensional blocks $A$, $B$, $C$, and $D$ as follows
\begin{equation}
M ~=~ \begin{pmatrix} A&B\\C&D \end{pmatrix} ~\in~ \Sp{2g}\;.
\end{equation}
Then, $M$ acts on $\Omega$ as
\begin{equation}
\label{eq:SP4ZModuliTrafo}
\Omega ~\stackrel{M}{\longrightarrow}~ \left( A\, \Omega + B \right) \left( C\, \Omega + D \right)^{-1}\;.
\end{equation}
Note that $\pm M \in \Sp{2g}$ yield the same transformation eq.~\eqref{eq:SP4ZModuliTrafo} of 
$\Omega$.

In the following, we focus on $g=2$. In this case, the moduli are encoded in a symmetric 
$2 \times 2$ matrix $\Omega$ whose components are denoted as
\begin{equation}
\label{eq:OmegaMatrix}
\Omega ~=~ \begin{pmatrix}U & Z\\ Z & T\end{pmatrix}\;.
\end{equation}
As we will see explicitly in the following, $T$ and $U$ are two moduli associated with the modular 
group $\SL{2,\Z{}}_{T} \times \SL{2,\Z{}}_{U}$, while $Z$ is a new modulus that interrelates the 
two $\SL{2,\Z{}}$ factors.

\subsection{\boldmath Subgroups of the Siegel modular group \Sp{4} \unboldmath}

The Siegel modular group \Sp{4} contains two factors of $\SL{2,\Z{}}$, i.e.
\begin{equation}
M_{(\gamma_T,\gamma_U)} ~:=~
\begin{pmatrix}
a_U & 0   & b_U & 0\\
0   & a_T & 0   & b_T\\
c_U & 0   & d_U & 0\\
0   & c_T & 0   & d_T
\end{pmatrix} ~\in~ \Sp{4}\;,
\end{equation}
where $a_T d_T-b_T c_T=a_U d_U-b_U c_U=1$. Hence,
\begin{equation}
\gamma_T ~:=~ \begin{pmatrix}a_T & b_T\\ c_T & d_T\end{pmatrix} ~\in~ \SL{2,\Z{}}_{T} \quad\mathrm{and}\quad \gamma_U ~:=~ \begin{pmatrix}a_U & b_U\\ c_U & d_U\end{pmatrix} ~\in~ \SL{2,\Z{}}_{U}\;.
\end{equation}
Here, $\SL{2, \Z{}}$ denotes the modular group generated by
\begin{equation}
\mathrm{S} ~:=~ \begin{pmatrix}0 & 1\\ -1 & 0\end{pmatrix} \quad\mathrm{and}\quad \mathrm{T} ~:=~ \begin{pmatrix}1 & 1\\ 0 & 1\end{pmatrix}\;.
\end{equation}
In detail, $\SL{2,\Z{}}_{U}$ is contained in \Sp{4} because
\begin{equation}
M_{(\Id_2,\gamma_U)} = \begin{pmatrix}
a_U&0&b_U&0\\
0&1&0&0\\
c_U&0&d_U&0\\
0&0&0&1
\end{pmatrix} \in \Sp{4} \quad\mathrm{as\ long\ as}\quad \gamma_U = \begin{pmatrix}a_U & b_U\\ c_U & d_U\end{pmatrix} \in \SL{2,\Z{}}_{U}\;,
\end{equation} 
due to the defining condition $M_{(\Id_2,\gamma_U)}^{\mathrm{T}}J M_{(\Id_2,\gamma_U)} = J$ of 
\Sp{4}, see eq.~\eqref{eq:DefSp2gZ}. Then, we use eq.~\eqref{eq:SP4ZModuliTrafo} and find that the 
moduli transform as 
\begin{subequations}\label{eq:SP4ZSubgroupSL2Ztau1}
\begin{eqnarray}
T & \xyrightarrow{M_{(\Id_2,\gamma_U)}} & T - \dfrac{c_U\, Z^{2}}{c_U\, U + d_U}\;,\\
U & \xyrightarrow{M_{(\Id_2,\gamma_U)}} & \dfrac{a_U\, U + b_U}{c_U\, U + d_U} \;,\\
Z & \xyrightarrow{M_{(\Id_2,\gamma_U)}} & \dfrac{Z}{c_U\, U + d_U}\;.
\end{eqnarray}
\end{subequations}
Note that for $Z=0$ we see that $T$ and $Z$ are invariant under $\SL{2,\Z{}}_{U}$ modular 
transformations, while $U$ transforms as expected from $\SL{2,\Z{}}_{U}$. Similarly, we can embed 
$\SL{2,\Z{}}_{T}$ into \Sp{4} via
\begin{equation}
M_{(\gamma_T,\Id_2)} ~=~ \begin{pmatrix}
1 & 0   & 0 & 0\\
0 & a_T & 0 & b_T\\
0 & 0   & 1 & 0\\
0 & c_T & 0 & d_T
\end{pmatrix} ~\in~ \Sp{4} \quad\text{while}\quad \gamma_T ~=~ \begin{pmatrix}a_T & b_T\\ c_T & d_T\end{pmatrix} ~\in~ \SL{2,\Z{}}_{T}\;,
\end{equation} 
such that the moduli transform as 
\begin{subequations}\label{eq:SP4ZSubgroupSL2Ztau2}
\begin{eqnarray}
T & \xyrightarrow{M_{(\gamma_T, \Id_2)}} & \dfrac{a_T\, T + b_T}{c_T\, T + d_T}\;,\\
U & \xyrightarrow{M_{(\gamma_T, \Id_2)}} & U-\dfrac{c_T\, Z^{2}}{c_T\, T + d_T}\;,\\
Z & \xyrightarrow{M_{(\gamma_T, \Id_2)}} & \dfrac{Z}{c_T\, T + d_T}\;,
\end{eqnarray}
\end{subequations}
using eq.~\eqref{eq:SP4ZModuliTrafo}. Let us remark that the modular $\mathrm{S}^2$ transformations 
from $\SL{2,\Z{}}_{T}$ and $\SL{2,\Z{}}_{U}$ are related in \Sp{4}, i.e.\ 
$M_{(\mathrm{S}^2, \Id_2)} = -M_{(\Id_2, \mathrm{S}^2)}$ and the moduli transform as
\begin{subequations}
\begin{eqnarray}
T & \stackrel{M}{\longrightarrow} & T\;,\\
U & \stackrel{M}{\longrightarrow} & U\;,\\
Z & \stackrel{M}{\longrightarrow} & -Z\;,
\end{eqnarray}
\end{subequations}
for $M\in\{M_{(\mathrm{S}^2, \Id_2)}, M_{(\Id_2, \mathrm{S}^2)}\}$.

In addition, \Sp{4} contains a $\Z{2}$ mirror transformation 
\begin{equation}
M_\times ~:=~ \begin{pmatrix}
0&1&0&0\\
1&0&0&0\\
0&0&0&1\\
0&0&1&0
\end{pmatrix} ~\in~ \Sp{4} \quad\mathrm{with}\quad \left(M_\times\right)^2 ~=~ \Id_4\;.
\end{equation}
As the name suggests, a mirror transformation interchanges $T$ and $U$, i.e.\ using 
eq.~\eqref{eq:SP4ZModuliTrafo} one can verify easily that
\begin{subequations}\label{eq:SP4ZMirror}
\begin{eqnarray}
T & \stackrel{M_\times}{\longrightarrow} & U\;,\\
U & \stackrel{M_\times}{\longrightarrow} & T\;,\\
Z & \stackrel{M_\times}{\longrightarrow} & Z\;.
\end{eqnarray}
\end{subequations}
Finally, \Sp{4} contains elements $M(\Delta)$ with $\Delta\in\Z{}^2$. These elements are 
intrinsically tied to the modulus $Z$. They can be defined as
\begin{equation}
M(\Delta) ~:=~ \begin{pmatrix}
 1 & 0 & 0    &-\ell\\
 m & 1 &-\ell & 0\\
 0 & 0 & 1    &-m\\
 0 & 0 & 0    & 1
\end{pmatrix}~\in~ \Sp{4} \quad\mathrm{for}\quad \Delta ~:=~ \begin{pmatrix}\ell\\ m\end{pmatrix} ~\in~\Z{}^2\;.
\end{equation}
Then, eq.~\eqref{eq:SP4ZModuliTrafo} yields
\begin{subequations}\label{eq:WLShiftOfModuli}
\begin{eqnarray}
T & \xyrightarrow{M(\Delta)} & T + m \left(m\,U + 2\, Z - \ell\right)\;,\\
U & \xyrightarrow{M(\Delta)} & U\;,\\
Z & \xyrightarrow{M(\Delta)} & Z + m\, U - \ell\;.
\end{eqnarray}
\end{subequations}

\newpage
\section{\boldmath The origin of the \Sp{4} Siegel modular group from strings\unboldmath}
\label{sec:OriginOfSP4Z}

It is well-known that compactifications of heterotic string theory on tori (and toroidal orbifolds) 
are naturally described in the Narain formulation~\cite{Narain:1985jj,Narain:1986am,Narain:1986qm}. 
There, one considers $D$ right- and $D+16$ left-moving (bosonic) string modes 
$(y_\mathrm{R}, y_\mathrm{L})$ to be compactified as
\begin{equation}\label{eq:NarainBC}
Y ~\sim~ Y + E\,\hat{N}\;, \quad\mathrm{where}\quad Y ~:=~ \begin{pmatrix}y_\mathrm{R}\\y_\mathrm{L}\end{pmatrix}\;,
\end{equation}
i.e.\ on an auxiliary torus of dimension $2D+16$. The 16 extra left-moving degrees of freedom give 
rise to an $\E{8}\times\E{8}$ (or $\SO{32}$) gauge symmetry of the heterotic string. In more 
detail, the auxiliary torus corresponding to the identification~\eqref{eq:NarainBC} can be 
defined by the so-called Narain lattice
\begin{equation}
\Gamma ~:=~ \Bigg\{ E\,\hat{N} ~\big{|}~ \hat{N} = \begin{pmatrix}n\\m\\p\end{pmatrix} \in \Z{}^{2D+16}\Bigg\} 
\end{equation}
that is spanned by the Narain vielbein $E$, a matrix of dimension $(2D+16)\times (2D+16)$. Here, 
$n\in\Z{}^D$ gives the winding numbers, $m\in\Z{}^D$ the Kaluza--Klein numbers and $p\in\Z{}^{16}$ 
the gauge quantum numbers. As the one-loop partition function of the string worldsheet has to be 
modular invariant, the Narain lattice $\Gamma$ has to be an even, integer and self-dual lattice 
with a metric $\eta$ of signature $(D,D+16)$. This condition on $\Gamma$ holds if the Narain 
vielbein $E$ satisfies
\begin{equation}\label{eq:NarainMetric}
\hat\eta ~:=~ E^\mathrm{T}\eta\,E ~=~ \begin{pmatrix}0 & \Id_D & 0\\\Id_D & 0 & 0\\ 0 & 0 & g\end{pmatrix},\quad\text{where}\quad \eta ~:=~ \begin{pmatrix}-\Id_D & 0 & 0\\0 & \Id_D & 0\\ 0 & 0 & \Id_{16}\end{pmatrix}\;.
\end{equation}
Here, $g := \alpha_\mathrm{g}^\mathrm{T} \alpha_\mathrm{g}$ is the Cartan matrix of the 
$\E{8}\times\E{8}$ gauge symmetry and $\alpha_\mathrm{g}$ denotes a matrix whose columns are the 
simple roots of $\E{8}\times\E{8}$ (or in the case of an $\SO{32}$ gauge symmetry, 
$\alpha_\mathrm{g}$ is a basis of the $\mathrm{Spin}(32)/\Z{2}$ weight lattice).

It is convenient to define the so-called generalized metric of the Narain lattice in terms of the 
metric $G:=e^\mathrm{T}e$ (of the $D$-dimensional torus spanned by the geometrical vielbein $e$), 
the anti-symmetric $B$-field $B$ and the Wilson lines $A$,
\begin{equation}\label{eq:GeneralizedMetric}
\mathcal{H} ~:=~ E^\mathrm{T}E ~:=~ \begin{pmatrix}
\frac{1}{\alpha'}\left(G + \alpha' A^\mathrm{T} A + C^\mathrm{T} G^{-1} C\right) & 
-C^\mathrm{T} G^{-1} & 
(\Id_2 + C^\mathrm{T} G^{-1}) A^\mathrm{T} \alpha_\mathrm{g} \\
-G^{-1} C & \alpha' G^{-1} & - \alpha' G^{-1} A^\mathrm{T} \alpha_\mathrm{g}\\
\alpha_\mathrm{g}^\mathrm{T} A (\Id_2 + G^{-1} C) & 
-\alpha' \alpha_\mathrm{g}^\mathrm{T} A G^{-1} & 
\alpha_\mathrm{g}^\mathrm{T} \big(\Id_{16}+ \alpha' A G^{-1} A^\mathrm{T} \big) \alpha_\mathrm{g}
\end{pmatrix}\;,
\end{equation}
where $C := B + \frac{\alpha'}{2} A^\mathrm{T} A$ and we use conventions similar to those of 
refs.~\cite{GrootNibbelink:2017usl,GrootNibbelink:2020dib}, but replacing $C$ by $C^\mathrm{T}$ for 
later convenience. Note that due to eq.~\eqref{eq:NarainMetric}, the generalized metric $\mathcal{H}$ 
satisfies the condition
\begin{equation}
\left(\mathcal{H}\,\hat\eta^{-1}\right)^2 ~=~ \Id_{2D+16}\;.
\end{equation}

The outer automorphisms of the Narain lattice are given by ``rotational'' transformations
\begin{equation}
\mathrm{O}_{\hat\eta}(D,D+16,\Z{}) ~:=~ \big\langle ~\hat\Sigma~\big|~ \hat\Sigma ~\in~\mathrm{GL}(2D+16,\Z{}) 
\quad\mathrm{with}\quad \hat\Sigma^\mathrm{T}\hat\eta\,\hat\Sigma = \hat\eta~\big\rangle\;.
\end{equation}
This is the general modular group of a toroidal compactification of the heterotic string. Elements 
$\hat\Sigma$ of $\mathrm{O}_{\hat\eta}(D,D+16,\Z{})$ act on the Narain vielbein $E$ 
as~\cite{Baur:2019iai}
\begin{equation}\label{eq:TrafoOfE}
E ~\stackrel{\hat\Sigma}{\longrightarrow}~ E\, \hat\Sigma^{-1}\;,
\end{equation}
such that the Narain scalar product $\lambda_1^\mathrm{T}\eta\lambda_2$ is invariant for 
$\lambda_i\in\Gamma$, $i\in\{1,2\}$.

In the following we take $D=2$. Moreover, the (continuous) Wilson lines are chosen as 
$A_{i}=(a_i,-a_i,0^{14})^\mathrm{T}$ for $i\in\{1,2\}$, where $A_{i}$ denote the two columns of 
$A$. Thus, we allow for continuous Wilson lines $a_i$ in the direction of the simple root 
$(1,-1,0^{14})^\mathrm{T}$ of $\E{8}\times\E{8}$ (or $\SO{32}$) for both directions $e_i$ of the 
geometrical two-torus. Then, we define moduli $(T,U,Z)$ of the two-torus with $B$-field 
\begin{equation}
B ~:=~ \alpha' b\, \epsilon\;, \quad\mathrm{where}\quad \epsilon ~:=~ \begin{pmatrix}0&1\\-1&0\end{pmatrix}\;,
\end{equation}
and Wilson lines background fields $A$ as
\begin{subequations}\label{eq:TUZ}
\begin{eqnarray}
T & := & \frac{1}{\alpha'} \left(B_{12} + \I\, \sqrt{\det G}\right) + a_{1}\left(-a_{2} + U\,a_{1} \right)\;,\\
U & := & \frac{1}{G_{11}} \left(G_{12} + \I\, \sqrt{\det G}\right) ~=~ \frac{|e_2|}{|e_1|}\,\textrm{e}^{\I \phi}\;,\\
Z & := & -a_{2} + U\,a_{1}\;,\label{eq:TUZ-Z}
\end{eqnarray}
\end{subequations}
cf.\ ref.~\cite{Mayr:1995rx}. Moreover, $e_1$ and $e_2$ are the two columns of the geometrical 
vielbein $e$, and $\phi$ denotes the angle enclosed by them. Note that the continuous Wilson lines 
$a_1$ and $a_2$ not only yield a new ``Wilson line modulus'' called $Z$ but they also alter the 
definition of the K\"ahler modulus $T$. In contrast, the complex structure modulus $U$ remains 
unchanged in the presence of Wilson lines. 

In what follows, it will be important to compute the transformation of the moduli $(T,U,Z)$ under 
general modular transformations from $\mathrm{O}_{\hat\eta}(2,3,\Z{})\subset\mathrm{O}_{\hat\eta}(2,2+16,\Z{})$. 
To do so, the generalized metric $\mathcal{H}=E^\mathrm{T} E$ and eq.~\eqref{eq:TrafoOfE} can be 
used to obtain 
\begin{equation}\label{eq:TrafoOfGeneralizedMetric}
\mathcal{H}(T,U,Z) ~\stackrel{\hat\Sigma}{\longrightarrow}~ \hat\Sigma^{-\mathrm{T}} \mathcal{H}(T,U,Z) \hat\Sigma^{-1} ~=:~ \mathcal{H}(T',U',Z')\;,
\end{equation}
for a general modular transformation $\hat\Sigma \in \mathrm{O}_{\hat\eta}(2,3,\Z{})$.

\subsection{\boldmath Mapping between $\mathrm{O}_{\hat\eta}(2,3,\Z{})$ of the Narain lattice and $\Sp{4}$\unboldmath}
In this section, we discuss various subgroups and generators of 
$\mathrm{O}_{\hat\eta}(2,3,\Z{})\subset\mathrm{O}_{\hat\eta}(2,2+16,\Z{})$, derive their actions on 
the moduli $(T,U,Z)$ and compare them to the Siegel modular group $\Sp{4}$. By doing so, we will 
show explicitly that the Siegel modular group $\Sp{4}$ appears naturally in toroidal 
compactifications of the heterotic string, see also 
refs.~\cite{LopesCardoso:1994is,Bailin:1998yt,Malmendier:2014uka,Font:2016odl,Font:2020rsk}. The main results are 
summarized in table~\ref{tab:Mapping} at the end of this section.

\paragraph{Mirror transformation.} We define a so-called mirror transformation
\begin{equation}\label{eq:Mirror}
\hat{M} ~:=~ \begin{pmatrix}0&0&-1&0&0\\0&1&0&0&0\\-1&0&0&0&0\\0&0&0&1&0\\0&0&0&0&\Id_{16}\end{pmatrix} ~\in~ \mathrm{O}_{\hat\eta}(2,2+16,\Z{})\;,
\end{equation}
where we have to change the conventions compared to refs.~\cite{Baur:2019kwi,Baur:2019iai} 
due to the presence of Wilson lines and the resulting changes in the generalized metric 
eq.~\eqref{eq:GeneralizedMetric}. Using eq.~\eqref{eq:TrafoOfGeneralizedMetric} we obtain
\begin{equation}
T ~\leftrightarrow~ U \quad\mathrm{and}\quad Z ~\leftrightarrow~ Z\;,
\end{equation}
as expected for a mirror transformation, see eq.~\eqref{eq:SP4ZMirror} for the corresponding case 
in $\Sp{4}$.

\paragraph{Modular group of the complex structure modulus.} The general modular group 
$\mathrm{O}_{\hat\eta}(2,2+16,\Z{})$ contains a modular group $\SL{2,\Z{}}_U$ associated with the 
complex structure modulus $U$. It can be generated by
\begin{equation}
\hat{C}_\mathrm{S} ~:=~ \begin{pmatrix}0&-1&0&0&0\\1&0&0&0&0\\0&0&0&-1&0\\0&0&1&0&0\\0&0&0&0&\Id_{16}\end{pmatrix} \quad\mathrm{and}\quad
\hat{C}_\mathrm{T} ~:=~ \begin{pmatrix}1&-1&0&0&0\\0&1&0&0&0\\0&0&1&0&0\\0&0&1&1&0\\0&0&0&0&\Id_{16}\end{pmatrix}\;.
\end{equation}
Then, we use eq.~\eqref{eq:TrafoOfGeneralizedMetric} in order to verify the $\Sp{4}$ transformations 
of the moduli $(T,U,Z)$ given in eq.~\eqref{eq:SP4ZSubgroupSL2Ztau1}.

\paragraph{Modular group of the K\"ahler modulus.}
In addition to $\SL{2,\Z{}}_U$, due to mirror symmetry eq.~\eqref{eq:Mirror} there exists a modular 
group $\SL{2,\Z{}}_T$ associated with the K\"ahler modulus $T$. It can be defined by
\begin{subequations}
\begin{eqnarray}
\hat{K}_\mathrm{S} & := & \hat{M}\, \hat{C}_\mathrm{S}\,\hat{M}^{-1} ~=~ \begin{pmatrix}0&0&0&1&0\\0&0&-1&0&0\\0&1&0&0&0\\-1&0&0&0&0\\0&0&0&0&\Id_{16}\end{pmatrix} \quad\mathrm{and}\\
\hat{K}_\mathrm{T} & := & \hat{M}\, \hat{C}_\mathrm{T}\,\hat{M}^{-1} ~=~ \begin{pmatrix}1&0&0&0&0\\0&1&0&0&0\\0&1&1&0&0\\-1&0&0&1&0\\0&0&0&0&\Id_{16}\end{pmatrix}\;.
\end{eqnarray}
\end{subequations}
These transformations reproduce the $\Sp{4}$ transformations eq.~\eqref{eq:SP4ZSubgroupSL2Ztau2} of 
$(T,U,Z)$, as can be seen explicitly using eq.~\eqref{eq:TrafoOfGeneralizedMetric}.

\paragraph{Wilson line shifts.} Due to the 16 extra left-moving degrees of freedom of the heterotic 
string, the general modular group $\mathrm{O}_{\hat\eta}(2,2+16,\Z{})$ has additional elements 
called ``Wilson line shifts''. They are defined as
\begin{equation}
\hat{W}(\Delta A) ~:=~ \begin{pmatrix}\Id_2&0&0\\-\frac{1}{2}\Delta A^\mathrm{T} g\, \Delta A&\Id_2&\Delta A^\mathrm{T} g\\ -\Delta A&0&\Id_{16}\end{pmatrix} ~\in~ \mathrm{O}_{\hat\eta}(2,2+16,\Z{})\;,
\end{equation}
where $\Delta A$ is a $16 \times 2$-dimensional matrix with integer entries. Since $g$ is the 
Cartan matrix of an even lattice (of $\E{8}\times\E{8}$ or $\mathrm{Spin}(32)/\Z{2}$), the 
$2 \times 2$ matrix $\frac{1}{2}\Delta A^\mathrm{T} g\, \Delta A$ is integer. We focus on shifts 
$\Delta A$ in the directions of $a_1$ and $a_2$. Hence, we define
\begin{equation}
\W{\ell}{m} ~:=~ \hat{W}(\Delta A) \quad\mathrm{for}\quad \Delta A ~:=~ \begin{pmatrix}m & \ell\\0&0\\ \vdots & \vdots\\ 0 & 0\end{pmatrix}\;,
\end{equation}
for $\ell, m \in \Z{}$. By doing so, we will focus in what follows on a subgroup 
$\mathrm{O}_{\hat\eta}(2,3,\Z{})$ of $\mathrm{O}_{\hat\eta}(2,2+16,\Z{})$. Then, using the 
transformation~\eqref{eq:TrafoOfGeneralizedMetric} of the generalized metric, we obtain
\begin{equation}
a_1 ~\xyrightarrow{\W{\ell}{m}}~ a_1 + m\;,\quad a_2 ~\xyrightarrow{\W{\ell}{m}}~ a_2 + \ell \quad\mathrm{and}\quad 
b ~\xyrightarrow{\W{\ell}{m}}~ b + a_1\, \ell - a_2\, m\;,
\end{equation}
while the metric $G$ is invariant. Translated to the moduli $(T,U,Z)$ defined in 
eq.~\eqref{eq:TUZ}, this reproduces the $\Sp{4}$ transformations given in 
eq.~\eqref{eq:WLShiftOfModuli}.

\paragraph{\boldmath \CP-like transformation.\unboldmath} Finally, as discussed in 
ref.~\cite{Baur:2019iai}, a \CP-like generator has to act not only on the $(2+2)$-dimensional 
Narain coordinates of the geometrical two-torus but also on the 16 extra left-moving degrees of 
freedom, i.e.
\begin{equation}
\hat{\Sigma}_* ~:=~ \begin{pmatrix}-1&0&0&0&0\\0&1&0&0&0\\0&0&-1&0&0\\0&0&0&1&0\\0&0&0&0&-\Id_{16}\end{pmatrix} ~\in~ \mathrm{O}_{\hat\eta}(2,2+16,\Z{})\;.
\end{equation}
Applying eq.~\eqref{eq:TrafoOfGeneralizedMetric} to $\hat{\Sigma}_*$ gives rise to a \CP-like 
transformation
\begin{equation}\label{eq:O23ZOnTUZ}
T ~\stackrel{\hat{\Sigma}_*}{\longrightarrow}~ -\bar{T}\;,\quad U ~\stackrel{\hat{\Sigma}_*}{\longrightarrow}~ -\bar{U}\quad\mathrm{and}\quad Z ~\stackrel{\hat{\Sigma}_*}{\longrightarrow}~ -\bar{Z}
\end{equation}
of the moduli. This string result on \CP can also be understood from a bottom-up perspective as we 
will see in section~\ref{sec:GSp}.

As a remark, there exist further $\mathrm{O}_{\hat\eta}(2,2+16,\Z{})$ transformations not present 
in $\Sp{4}$: One can perform Weyl reflections in the 16-dimensional lattice of $\E{8}\times\E{8}$ 
(or $\mathrm{Spin}(32)/\Z{2}$), see for example $\hat{M}_\mathrm{W}(\Delta W)$ in 
ref.~\cite{GrootNibbelink:2017usl}. 

\begin{table}[t!]
	\centering
	\begin{tabular}{c|c|c|l}
		\toprule
		symmetry          & $\Sp{4}$                  & $\mathrm{O}_{\hat\eta}(2,3,\Z{})$ & transformation of moduli\\
		\hline
		\multirow{6}{*}{$\SL{2,\Z{}}_T$}   & \multirow{3}{*}{$M_{(\mathrm{S},\Id_2)}$} & \multirow{3}{*}{$\hat{K}_\mathrm{S}$}                 & $T \rightarrow -\frac{1}{T}$\\
		&                           &                                      & $U \rightarrow U-\frac{Z^2}{T}$ \\
		&                           &                                      & $Z \rightarrow -\frac{Z}{T}$\\
		\cline{2-4}
		& \multirow{3}{*}{$M_{(\mathrm{T}, \Id_2)}$} & \multirow{3}{*}{$\hat{K}_\mathrm{T}$}               & $T \rightarrow T+1$ \\
		&                           &                                      & $U \rightarrow U$ \\
		&                           &                                      & $Z \rightarrow Z$ \\
		\hline
		\multirow{6}{*}{$\SL{2,\Z{}}_U$}   & \multirow{3}{*}{$M_{(\Id_2, \mathrm{S})}$} & \multirow{3}{*}{$\hat{C}_\mathrm{S}$}                & $T \rightarrow T-\frac{Z^2}{U}$\\
		&                           &                                      & $U \rightarrow -\frac{1}{U}$ \\
		&                           &                                      & $Z \rightarrow -\frac{Z}{U}$\\
		\cline{2-4}
		& \multirow{3}{*}{$M_{(\Id_2, \mathrm{T})}$} & \multirow{3}{*}{$\hat{C}_\mathrm{T}$}                & $T \rightarrow T$\\
		&                           &                                      & $U \rightarrow U+1$ \\
		&                           &                                      & $Z \rightarrow Z$ \\
		\hline
		\multirow{3}{*}{Mirror}            & \multirow{3}{*}{$M_\times$}              & \multirow{3}{*}{$\hat{M}$}                            & $T \rightarrow U$ \\
		&                           &                                      & $U \rightarrow T$ \\
		&                           &                                      & $Z \rightarrow Z$ \\
		\hline
		\multirow{3}{*}{Wilson line shift} & \multirow{3}{*}{$\M{\ell}{m}$}             & \multirow{3}{*}{$\W{\ell}{m}$}                        & $T \rightarrow T + m \left(m\,U + 2\, Z - \ell\right)$ \\
		&                           &                                      & $U \rightarrow U$ \\
		&                           &                                      & $Z \rightarrow Z + m\, U - \ell$\\
		\hline
		\multirow{3}{*}{\CP-like}          & \multirow{3}{*}{\parbox[c]{50pt}{\hfill$M_* \in$\hfill\hfill\\$~\mathrm{GSp}(4,\Z{})$}}                 & \multirow{3}{*}{$\hat{\Sigma}_*$}                     & $T \rightarrow -\bar{T}$ \\
		&     &                                      & $U \rightarrow -\bar{U}$ \\
		&                           &                                      & $Z \rightarrow -\bar{Z}$ \\
		\bottomrule
	\end{tabular}
	\caption{We list the generators of the Siegel modular group \Sp{4} and their corresponding elements 
		in the subgroup $\mathrm{O}_{\hat\eta}(2,3,\Z{})$ of the general modular group 
		$\mathrm{O}_{\hat\eta}(2,2+16,\Z{})$ constructed explicitly in section~\ref{sec:OriginOfSP4Z} in 
		the Narain formulation of the heterotic string. In the last column we list the transformation of 
		the moduli, computed in two ways: i) using eq.~\eqref{eq:SP4ZModuliTrafo} for \Sp{4}, and 
		independently ii) using eq.~\eqref{eq:TrafoOfGeneralizedMetric} for 
		$\mathrm{O}_{\hat\eta}(2,3,\Z{})$. The \CP-like transformation $M_*$ will be 
		defined in section~\ref{sec:GSp}.}
	\label{tab:Mapping}
\end{table}

\section{\boldmath \CP as an outer automorphism of \Sp{4} \unboldmath}
\label{sec:GSp}

We have seen in the previous section that a \CP-like transformation appears naturally in (toroidal) 
string compactifications. As we shall see in this section in a bottom-up discussion, this 
transformation does not belong to \Sp{4} but corresponds to an outer automorphism of \Sp{4} that, 
once included, enhances \Sp{4} to the general symplectic group $\mathrm{GSp}(4,\Z{})$.

We define a transformation
\begin{equation}\label{eq:SP4ZActionOfMstar}
M ~\stackrel{M_*}{\longrightarrow}~ M' ~:=~ M_*^{-1} M\, M_* \quad\mathrm{for\ all}\quad M ~\in~ \Sp{4}\;,
\end{equation}
where $M_*$ is given by
\begin{equation}\label{eq:GSp}
M_* ~:=~ \begin{pmatrix}
-1 & 0 & 0 & 0\\
 0 &-1 & 0 & 0\\
 0 & 0 & 1 & 0\\
 0 & 0 & 0 & 1
\end{pmatrix} \quad\mathrm{satisfying}\quad M_*^{\mathrm{T}}J\, M_* ~=~ -J\;.
\end{equation}
Hence, $M_*\not\in\Sp{4}$. Rather it lies in the general symplectic group 
\begin{equation}
\mathrm{GSp}(4,\Z{}) ~:=~ \left\lbrace M ~\in~ \Z{}^{4\times 4} ~\vert~ M^{\mathrm{T}}J\, M ~=~ \pm J \right\rbrace\;.
\end{equation}
Then, it is easy to see that $M'$ defined in eq.~\eqref{eq:SP4ZActionOfMstar} is an element from 
\Sp{4} for all $M \in \Sp{4}$, i.e.
\begin{equation}
\left(M'\right)^\mathrm{T} J\, M' ~=~ M_*^\mathrm{T} M^\mathrm{T} \underbrace{M_*^{-\mathrm{T}} J\, M_*^{-1}}_{=~-J} M\, M_* ~=~ - M_*^\mathrm{T} \underbrace{M^\mathrm{T} J\, M}_{=~J}\, M_* ~=~ +J\;.
\end{equation}
Hence, eq.~\eqref{eq:SP4ZActionOfMstar} defines an automorphism of \Sp{4}. It is outer
because $M_*\not\in\Sp{4}$, as seen in eq.~\eqref{eq:GSp}.

In order to see the physical meaning of $M_*$, we apply eq.~\eqref{eq:SP4ZActionOfMstar} to various 
elements of \Sp{4}:
\begin{subequations}\label{eq:SP4ZCP}
\begin{eqnarray}
M_{(\mathrm{S}, \Id_2)} & \stackrel{M_*}{\longrightarrow} & M_*^{-1} M_{(\mathrm{S}, \Id_2)}\, M_* ~=~ \left(M_{(\mathrm{S}, \Id_2)}\right)^{-1}\;, \\
M_{(\mathrm{T}, \Id_2)} & \stackrel{M_*}{\longrightarrow} & M_*^{-1} M_{(\mathrm{T}, \Id_2)}\, M_* ~=~ \left(M_{(\mathrm{T}, \Id_2)}\right)^{-1}\;, \\
M_{(\Id_2, \mathrm{S})} & \stackrel{M_*}{\longrightarrow} & M_*^{-1} M_{(\Id_2, \mathrm{S})}\, M_* ~=~ \left(M_{(\Id_2, \mathrm{S})}\right)^{-1}\;, \\
M_{(\Id_2, \mathrm{T})} & \stackrel{M_*}{\longrightarrow} & M_*^{-1} M_{(\Id_2, \mathrm{T})}\, M_* ~=~ \left(M_{(\Id_2, \mathrm{T})}\right)^{-1}\;, \\
M_\times                & \stackrel{M_*}{\longrightarrow} & M_*^{-1} M_\times\, M_*                ~=~ \left(M_\times\right)^{-1} ~=~ M_\times\;,\\
\M{\ell}{m}             & \stackrel{M_*}{\longrightarrow} & M_*^{-1} \M{\ell}{m}\, M_*             ~=~ \M{-\ell}{m}\;.\label{eq:SP4ZCPf}
\end{eqnarray}
\end{subequations}
Let us analyze eq.~\eqref{eq:SP4ZCPf} in more detail: For each choice of $\ell,m$, one can find 
an $M\in\Sp{4}$, such that $M\M{-\ell}{m}M^{-1}=\M{-\ell}{-m}$, which implies that $\M{\ell}{m}$ is 
mapped by $M_*$ to the conjugacy class of its inverse $\M{-\ell}{-m}$. Motivated by 
eqs.~\eqref{eq:SP4ZCP}, we consider $M_*$ a \CP-like transformation, see 
refs.~\cite{Holthausen:2012dk,Chen:2014tpa}. Indeed, as explained in appendix~\ref{app:Genusg}, the 
action of $\mathrm{GSp}(4,\Z{})$ on $\Omega$ can be defined in analogy to the action of 
$\mathrm{GL}(2,\Z{})$ on one modulus, cf.\ ref.~\cite{Dent:2001cc,Baur:2019kwi,Novichkov:2019sqv} 
and ref.~\cite{Ishiguro:2020nuf}. Explicitly, for an element of the general symplectic group
\begin{equation}
M ~=~ \begin{pmatrix}A & B \\ C & D\end{pmatrix} ~\in~ \mathrm{GSp}(4,\Z{})
\end{equation}
we find the transformation rules
\begin{subequations}
\begin{eqnarray}
\label{eq:SP4ZModuliCPTrafo}
\Omega & \stackrel{M}{\longrightarrow} & \left( A\, \bar{\Omega} + B \right)   \left( C\, \bar{\Omega} + D \right)^{-1} \qquad   \mathrm{if}\quad M^{\mathrm{T}}J\, M = -J\;,\\
\Omega & \stackrel{M}{\longrightarrow} & \left( A\, \Omega       + B \right)\, \left( C\, \Omega       + D \right)^{-1} \,\qquad \mathrm{if}\quad M^{\mathrm{T}}J\, M = +J\;,
\end{eqnarray}
\end{subequations}
where $\bar{\Omega}$ denotes the complex conjugate of $\Omega$. Consequently, the moduli transform 
under $M_*$ as
\begin{equation}\label{eq:MstarOnTUZ}
T ~\stackrel{M_{*}}{\longrightarrow}~ -\bar{T}\;\;,\;\; U ~\stackrel{M_{*}}{\longrightarrow}~ -\bar{U}\;\;,\;\; Z ~\stackrel{M_{*}}{\longrightarrow}~ -\bar{Z}\;,
\end{equation}
which confirms our expectation for a $\CP$-like transformation.

\section{Conclusions and Outlook}
\label{sec:conclusions}

The potential (traditional \emph{and} modular) flavor symmetries of string theory compactifications 
are determined through the outer automorphisms of the Narain lattice. For the heterotic string, the 
modular symmetries are a subgroup of $\mathrm{O}_{\hat\eta}(D,D+16,\Z{})$, where $D$ is the 
dimension of the relevant compact space, i.e.\ $D\leq 6$. As a starting point, we have concentrated 
in this paper on a $D=2$ sublattice of compact six-dimensional space. Apart from the K\"ahler and 
complex structure moduli $T$ and $U$, we include a Wilson line modulus $Z$ and arrive at the 
modular symmetry group $\mathrm{O}_{\hat\eta}(2,3,\Z{})$. We show that this group is closely 
related to the Siegel modular group \Sp{4}, which has been studied intensively in the mathematical 
literature. The (complex) three-dimensional moduli space of \Sp{4} can be visualized through an 
auxiliary Riemann surface of genus 2 (see figure~\ref{fig:tori}). Our top-down construction allows 
for a physical interpretation of the recent bottom-up discussion of ref.~\cite{Ding:2020zxw}: Their 
``third'' modulus $\tau_3$ (apart from $\tau_1=U$ and $\tau_2=T$) can be understood as a Wilson 
line modulus $Z$ of compactified (heterotic) string theory. Furthermore, we have shown in a general 
study that, in addition to modular symmetries, there is a natural appearance of a \CP-like 
transformation predicted from the group $\mathrm{O}_{\hat\eta}(2,3,\Z{})$ in string theory. As 
discussed in section~\ref{sec:GSp}, from a bottom-up perspective, this \CP-like transformation can 
be understood as an outer automorphism of the Siegel modular group extending it to 
$\mathrm{GSp}(4,\Z{})$.

Beyond these results, an important open task is to make contact with realistic models of ``flavor'' 
including chiral matter. With this purpose, it is necessary to alter the $\mathbbm{T}^2$ toroidal 
compactification by a $\Z{K}$ orbifolding, i.e.\ $\mathbbm{T}^2/\Z{K}$. In string theory, this 
orbifolding results in the appearance of twisted strings, which in general give rise to chiral 
matter. Moreover, it is a remarkable fact that modular symmetries act nontrivially on twisted 
strings, such that twisted strings build nontrivial representations of a \emph{finite} modular flavor 
group. Thus, the orbifolding in string theory is instrumental to obtain chiral matter that exhibits 
finite modular flavor symmetries. This general mechanism of string theory has been discussed in 
detail for the $\mathbbm{T}^2/\Z{3}$ orbifold without Wilson lines: in this case, the modular 
symmetry $\SL{2,\Z{}}_T$ of the K\"ahler modulus $T$ acts as the \emph{finite} modular flavor symmetry 
$T'$ on the chiral matter from the twisted sectors of the 
orbifold~\cite{Baur:2019kwi,Baur:2019iai,Nilles:2020kgo}.

As we have seen in this work explicitly, the modular symmetry $\mathrm{O}_{\hat\eta}(2,3,\Z{})$ of 
a toroidal compactification of string theory \emph{with} Wilson lines corresponds to the Siegel 
modular group \Sp{4}. Hence, a $\Z{K}$ orbifolding can in general give rise to a \emph{finite} 
Siegel modular flavor group $\Gamma_{g,n}$ (here, of genus $g=2$), where chiral matter arises from 
the twisted sectors of the orbifold and builds nontrivial representations of $\Gamma_{2,n}$. For 
the \Z{2} orbifold we have $n=2$ and $\Gamma_{2,2}$ is isomorphic to $S_6$, the permutation group 
of six elements, see also ref.~\cite{Ding:2020zxw}. This $S_6$ includes the finite modular group 
$S_3\times S_3$ as well as mirror symmetry, as obtained in the string theory discussion of 
ref.~\cite{Baur:2020jwc}, where only the moduli $T$ and $U$ associated with 
$\SL{2,\Z{}}_T \times \SL{2,\Z{}}_U$ had been considered and the Wilson line modulus was set to 
$Z=0$. This indicates the path how to generalize to the case $Z \neq 0$ in string theory. In 
general, a $\Z{K}$ orbifolding can break the Siegel modular group \Sp{4} by discrete Wilson lines: 
The geometrical $\Z{K}$ rotation that acts on the two-torus has to be embedded into the 16 degrees 
of freedom of the gauge symmetry due to worldsheet modular invariance of the string partition 
function. It is known that a shift embedding yields discrete Wilson lines~\cite{Ibanez:1986tp}, 
such that the Wilson line modulus $Z$ is frozen at some discrete value. In this case, the Siegel 
modular group \Sp{4} is broken by the fixed Wilson line modulus. For a $\Z{K}$ orbifold with 
$K \neq 2$ the unbroken subgroup from \Sp{4} is at least the modular group $\SL{2,\Z{}}_T$ of the 
K\"ahler modulus $T$, while for $K=2$ one finds at least $\SL{2,\Z{}}_T \times \SL{2,\Z{}}_U$ 
combined with a mirror symmetry that interchanges $T$ and $U$, see ref.~\cite{Baur:2020jwc}. On the 
other hand, a rotational embedding into the 16 gauge degrees of freedom gives rise to continuous 
Wilson lines~\cite{Ibanez:1987xa,Forste:2005rs}, where the Wilson line modulus $Z$ remains as a 
free modulus. Hence, one expects that a two-dimensional \Z{2} orbifold with rotational embedding 
yields the full \Sp{4} Siegel modular group, where chiral matter from the twisted sector transforms 
in representations of the \emph{finite} Siegel modular flavor group $\Gamma_{2,2}\cong S_6$. A full 
discussion of the symmetries of the \Z{2} orbifold, including \CP, will be subject of a future 
publication~\cite{Baur:2021pr}.

\section*{Acknowledgments}
A.B. and P.V. are supported by the Deutsche Forschungsgemeinschaft (SFB1258).
The work of S.R.-S.\ was partly supported by CONACyT grant F-252167.

\appendix

\section{\boldmath Symplectic groups \Sp{2g} and modular transformations\unboldmath}
\label{app:Genusg}

\begin{figure}[t!]
 \centering
    \includegraphics{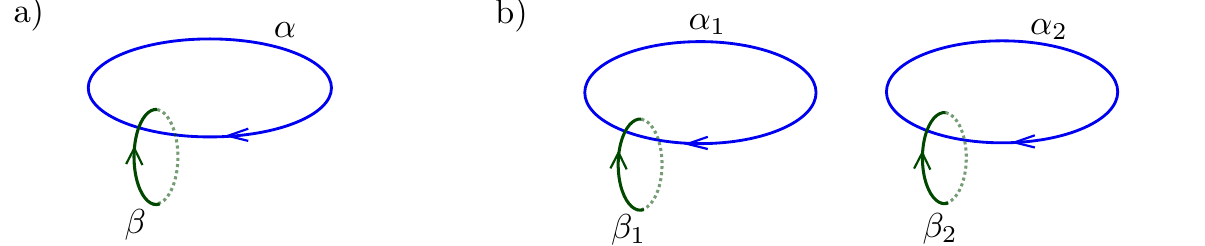}
    \caption{a) A $\mathcal T_1=\mathbbm T^2$ torus with the two basis 1-cycles, $\alpha$ and $\beta$.
    Its modular symmetry group is $\Sp{2}\cong\SL{2,\Z{}}$.
    b) A compact Riemann surface of genus 2 $\mathcal T_2$ and its four basis 1-cycles 
    $(\beta_1,\beta_2,\alpha_1,\alpha_2)^\mathrm{T}$. The Siegel modular group \Sp{4} is the modular
    symmetry group of $\mathcal T_2$. As discussed in ref.~\cite{Mayr:1995rx}, setting the Wilson 
    line modulus $Z$ defined in eq.~\eqref{eq:TUZ-Z} to $Z=0$ splits the genus 2 surface into two 
    separated two-tori. Note that these auxiliary surfaces must not be mistaken as compactification 
    spaces.}
    \label{fig:tori}
\end{figure}

Let us review some aspects of the symplectic group \Sp{2g} and its relation to modular 
transformations (see e.g.~\cite{Brownstein:1992,Perna:2016} for further details). The symplectic 
group \Sp{2g} can be defined by considering an auxiliary genus-$g$ Riemann surface $\mathcal T_g$ 
and its symmetries as follows: The genus-$g$ surface has $2g$ nontrivial 1-cycles denoted by 
$(\beta_i,\alpha_j)$ for $i,j\in\{1,\ldots,g\}$, see figure~\ref{fig:tori} for the cases $g=1$ and 
$g=2$. These cycles form the canonical basis of the homology group 
$H_1(\mathcal T_g,\Z{})\cong \Z{}^{2g}$. The holomorphic 1-forms $\omega_i$ build the dual 
cohomology basis, which one can choose to satisfy $\int_{\alpha_j}\omega_i=\delta_{ij}$ and 
$\int_{\beta_j}\omega_i = \int_{\beta_i}\omega_j$. In these terms, the skew-symmetric form $J$ in 
eq.~\eqref{eq:DefSp2gZ} is interpreted as the intersection numbers 
$\s{\beta}{\alpha}\cap\s{\beta}{\alpha}$ of the $2g$-dimensional vectors of 1-cycles 
$\s{\beta}{\alpha}=(\beta_1,\dots,\beta_g,\alpha_1,\dots,\alpha_g)^\mathrm{T}$, such that 
$\alpha_i\cap\alpha_j = \beta_i\cap\beta_j = 0$ and 
$-(\alpha_i\cap\beta_j)=\beta_i\cap\alpha_j=\delta_{ij}$. Now, one transforms the 1-cycles 
$\s{\beta}{\alpha}$ 
\begin{equation}\label{eq:Trafoalphabeta}
\s{\beta}{\alpha} ~\stackrel{M}{\longrightarrow}~ \s{\beta'}{\alpha'} ~:=~ \left(\begin{smallmatrix}A&B\\C&D\end{smallmatrix}\right) \s{\beta}{\alpha} ~=~ \s{A\,\beta+B\,\alpha}{C\,\beta+D\,\alpha}\quad\mathrm{for}\quad M~=~ \left(\begin{smallmatrix}A&B\\C&D\end{smallmatrix}\right) ~\in~ \Z{}^{g \times g}\;.
\end{equation}
The new 1-cycles $\s{\beta'}{\alpha'}$ also form a basis of $H_1(\mathcal T_g,\Z{})$ if 
$M\in\mathrm{GL}(2g,\Z{})$. Moreover, we have to require that the intersection numbers and, hence, 
$J$ be invariant under the transformation~\eqref{eq:Trafoalphabeta}. This amounts to demanding that 
$M\,J\,M^\mathrm{T}=J$. By taking the inverse transpose of this equation we get 
$M^{-\mathrm{T}}\,J^{-\mathrm{T}}\,M^{-1}=J^{-\mathrm{T}}$. Then, using $J^{-\mathrm{T}}=J$ we 
obtain the condition $M^\mathrm{T}\,J\,M=J$, i.e.\ $M\in\Sp{2g}$.

A consequence of the Torelli theorem for Riemann surfaces is that the genus-$g$ surface 
$\mathcal T_g$ is determined by the complex $g$-dimensional torus, which can be defined as the 
quotient of $\mathbbm C^g$ divided by a complex lattice. 
This lattice is given by the columns of the $g\x2g$ period matrix of $\mathcal T_g$,
\begin{equation}
\label{eq:periodM}
 \Pi_g ~:=~\begin{pmatrix}
 \int_{\alpha_1} \omega_1 & \dots & \int_{\alpha_1} \omega_g & \int_{\beta_1} \omega_1 & \dots & \int_{\beta_1} \omega_g\\
 \vdots                   &       & \vdots                   & \vdots                  &       & \vdots\\
 \int_{\alpha_g} \omega_1 & \dots & \int_{\alpha_g} \omega_g & \int_{\beta_g} \omega_1 & \dots & \int_{\beta_g} \omega_g 
 \end{pmatrix}\,.
 \end{equation}
By choosing a basis in which $\int_{\alpha_j}\omega_i=\delta_{ij}$, one can always rewrite $\Pi_g$, 
such that
\begin{equation}
 \Pi_g~=~ (\Id_g, \Omega)\;,
\end{equation}
where we have defined the $g\x g$ complex modulus matrix $\Omega$, such that 
$\Omega^\mathrm{T}=\Omega$. Clearly, the transformations $\s{\beta}{\alpha}\to M \s{\beta}{\alpha}$ 
induce transformations on the modulus matrix $\Omega$ in eq.~\eqref{eq:periodM}.
Restricting further to $\im\Omega>0$, we arrive at the modular space of the genus-$g$ compact 
surface $\mathcal T_g$,
\begin{equation}
\mathbbm H_g = \left\{\Omega\in\mathbbm{C}^{g\x g}~|~ \Omega^\mathrm{T}=\Omega, \im\Omega>0 \right\}\;.
\end{equation}

Consider the $g=1$ case. We observe that $M$ are $2\x2$ integer matrices with unit determinant, 
i.e.\ they describe the modular group $\Sp{2}\cong\SL{2,\Z{}}$ of a $\mathbbm T^2$ torus. Given the 
holomorphic 1-form $\omega=\dd z$ and the nontrivial 1-cycles $\alpha$ and $\beta$, shown in 
figure~\ref{fig:tori}a), the period matrix of $\mathbbm T^2$ is given by
\begin{equation}
 \Pi_1 ~=~ \left(\textstyle\int_\alpha \omega, \int_\beta \omega \right) ~=~ (1,\tau)\,,\qquad
 \tau\in\mathbbm C\,,~\im\tau>0\,.
\end{equation}
The last equation arises from the choice $\int_\alpha\omega = 1$ and the definition of the modulus 
$\tau:=\int_\beta\omega$. We now let the $\SL{2,\Z{}}$ element $M=\left(\begin{smallmatrix}a&b\\c&d\end{smallmatrix}\right)$ 
act on the 1-cycle vector $(\beta,\alpha)^\mathrm{T}$. This implies that the period matrix 
transforms as
\begin{equation}
\Pi_1 ~\to~ \Pi_1' = \left(\textstyle\int_{c\beta+d\alpha} \omega', \int_{a\beta+b\alpha} \omega' \right)
=\left(c\textstyle\int_\beta \omega' + d\int_\alpha \omega', a\int_\beta \omega'+b\int_\alpha \omega' \right)\,.
\end{equation}
By demanding that the holomorphic 1-form transforms under $M$ as $\omega'=\omega(c\tau+d)^{-1}$, we 
normalize the transformed period matrix, which then becomes
\begin{equation}
\Pi_1' ~=~ \left(1, (a\tau+b)(c\tau+d)^{-1}\right)\,.
\end{equation}
This allows us to identify the standard modular transformation 
$\tau\to (a\tau+b)(c\tau+d)^{-1}$.

The same discussion can be conducted for the more interesting case $g=2$, which leads to the Siegel 
modular group $\Sp4$. The Riemann surface $\mathcal T_2$ has the holomorphic 1-form basis 
$(\omega_1,\omega_2)$ and the nontrivial 1-cycles $\s{\beta}{\alpha}=(\beta_1,\beta_2,\alpha_1,\alpha_2)^\mathrm{T}$, 
illustrated in figure~\ref{fig:tori}b). Thus, its $2\x4$ period matrix reads
\begin{equation}
\Pi_2 ~=~ \begin{pmatrix}
\textstyle\int_{\alpha_1}\omega_1 & \int_{\alpha_1}\omega_2 & \int_{\beta_1}\omega_1 & \int_{\beta_1}\omega_2\\
\int_{\alpha_2}\omega_1 & \int_{\alpha_2}\omega_2 & \int_{\beta_2}\omega_1 & \int_{\beta_2}\omega_2
\end{pmatrix} ~=~ (\Id_2,\,\Omega)\,,
\end{equation}
where in the last relation we have chosen $\int_{\alpha_j}\omega_i=\delta_{ij}$ and defined the 
modular matrix $\Omega$, as given in eq.~\eqref{eq:OmegaMatrix}, satisfying 
$\Omega=\Omega^\mathrm{T}$ and $\im\Omega>0$. Next, we perform an \Sp{4} transformation 
$M\s{\beta}{\alpha}$, where $M=\left(\begin{smallmatrix}A&B\\C&D\end{smallmatrix}\right)$ and 
$A,B,C,D$ are $2\x2$ integer matrices. Although the expression for the transformed period matrix 
$\Pi_2'$ is more complicated than in the case $g=1$, one can readily show that, by demanding that 
\Sp{4} transformations on the 1-forms act as 
$(\omega_1',\omega_2')=(\omega_1,\omega_2)(C\Omega + D)^{-1}$, one arrives at
\begin{equation}
 \Pi_2~\to~\Pi_2'~=~\left(\Id_2,\,(A\Omega+B)(C\Omega + D)^{-1}\right)\;.
\end{equation}
We find thus that \Sp{4} transformations act on the modular matrix $\Omega$ as 
\begin{equation}
\label{eq:Sp4ZTrafoOfOmega}
\Omega ~\stackrel{M}{\longrightarrow}~  \left(A\Omega+B\right)\left(C\Omega + D\right)^{-1}\;.
\end{equation}

We can also consider a $\mathrm{GSp}(4,\Z{})$ transformation, where we are interested in the 
transformation of $\Omega$ under those $\tilde{M}\in\mathrm{GSp}(4,\Z{})$ with 
$\tilde{M}^\mathrm{T}J \tilde{M} = -J$ (since for the case with $M^\mathrm{T}J M = +J$ the result 
is already given in eq.~\eqref{eq:Sp4ZTrafoOfOmega}). In section~\ref{sec:GSp}, we have defined a 
special element $M_* \in \mathrm{GSp}(4,\Z{})$ with $M_*^\mathrm{T}J M_* = -J$. Then, the combined 
transformation $M := \tilde{M}\,M_*$ satisfies $M^\mathrm{T}J M = +J$, so $M \in \Sp{4}$. Since we 
know the transformation of $\Omega$ for $M \in \Sp{4}$, we only need to know the transformation of 
$\Omega$ for our special element $M_*$ in order to understand the general case with 
$\tilde{M} = MM_*$. Under $M_*$ the 1-cycles $\s{\beta}{\alpha}$ transform as
\begin{equation}\label{eq:TrafoalphabetaMstar}
\s{\beta}{\alpha} ~\stackrel{M_*}{\longrightarrow}~ \left(\begin{smallmatrix}-\Id_2&0\\0&\Id_2\end{smallmatrix}\right) \s{\beta}{\alpha} ~=~ \s{-\beta}{\alpha}\;,
\end{equation}
cf.\ eq.~\eqref{eq:Trafoalphabeta}. As can be seen in figure~\ref{fig:tori}, this 
transformation corresponds to a geometrical mirror transformation at a horizontal plane. By 
choosing appropriate complex coordinates on the surface $\mathcal T_2$ (i.e.\ in each chart) this 
mirror transformation acts by complex conjugation. Consequently, it is conceivable that $M_*$ has 
to map the 1-forms $\omega_i$ to $\bar{\omega}_i$. For the period matrix, this amounts to
\begin{equation}
\Pi_2 ~=~ 
\begin{pmatrix}
\Id_2, \Omega
\end{pmatrix}~\stackrel{M_*}{\longrightarrow}~ 
\begin{pmatrix}
\textstyle\int_{\alpha_1}\bar\omega_1 & \int_{\alpha_1}\bar\omega_2 & \int_{-\beta_1}\bar\omega_1 & \int_{-\beta_1}\bar\omega_2\\
\int_{\alpha_2}\bar\omega_1 & \int_{\alpha_2}\bar\omega_2 & \int_{-\beta_2}\bar\omega_1 & \int_{-\beta_2}\bar\omega_2
\end{pmatrix} ~=~
\begin{pmatrix}
\Id_2, -\bar\Omega
\end{pmatrix}\;.
\end{equation}
This proves eq.~\eqref{eq:MstarOnTUZ} that we have also found independently in the string setup, 
see eq.~\eqref{eq:O23ZOnTUZ}. Furthermore, this discussion can be generalized easily to the case 
$\mathrm{GSp}(2g,\Z{})$ for $\mathcal T_g$ with general genus $g$. This is particularly easy for 
$g=1$, where the basis 1-form is $\omega=\dd z$. Considering the $\CP$-like transformation 
$\left(\begin{smallmatrix}-1&0\\0&1\end{smallmatrix}\right)$ (using e.g.\ eqs.~(66) and~(144) of 
ref.~\cite{Nilles:2020gvu}) we get 
$(\im\dd z,\re\dd z)\stackrel{M_*}{\longrightarrow}(-\im\dd z,\re\dd z)$. Consequently, we see that 
$\dd z\stackrel{M_*}{\longrightarrow}\dd\bar z$. It then follows for the period matrix that 
$\Pi_1=(1,\tau)\stackrel{M_*}{\longrightarrow} (\int_\alpha\bar\omega,\int_{-\beta}\bar\omega)=(1,-\bar\tau)$,
choosing $\tau=\int_\beta\omega$ and $\int_\alpha\omega=1$, as before. This
confirms the well-known \CP-like transformation of the modulus 
$\tau$~\cite{Dent:2001cc,Baur:2019kwi,Novichkov:2019sqv}, which
promotes the modular symmetry $\Sp{2}\cong\SL{2,\Z{}}$ to 
$\mathrm{GSp}(2,\Z{})\cong\mathrm{GL}(2,\Z{})$.

\section{\boldmath Relations between elements of \Sp{4} \unboldmath}

In this appendix, we state several relations between elements of the Siegel modular group \Sp{4}. 
We have verified that they also hold in $\mathrm{O}_{\hat\eta}(2,2+16,\Z{})$ using the dictionary 
given in table~\ref{tab:Mapping}. This gives a further non-trivial proof that \Sp{4} and 
$\mathrm{O}_{\hat\eta}(2,2+16,\Z{})$ are related.

Elements $M_{(\gamma_T, \gamma_U)}$ of $\SL{2,\Z{}}_{T} \times \SL{2,\Z{}}_{U} \subset \Sp{4}$ get 
multiplied as
\begin{equation}
M_{(\gamma_1, \gamma_2)}\,M_{(\delta_1, \delta_2)} ~=~ M_{(\gamma_1\,\delta_1, \gamma_2\,\delta_2)}\;,
\end{equation}
as might have been expected. Thus, the elements $M_{(\gamma_T, \gamma_U)}$ form a subgroup of \Sp{4}.

On the other hand, the set of elements $M(\Delta)\in\Sp{4}$ does not form a subgroup of \Sp{4} 
on its own as one can see from the relation
\begin{equation}
M(\Delta_1)\,M(\Delta_2) ~=~ M(\Delta_1+\Delta_2)\,\left(M_{(\mathrm{T}, \Id_2)}\right)^{\Delta_1^\mathrm{T}\epsilon\,\Delta_2}\;,
\end{equation}
for $\Delta_1$, $\Delta_2 \in \Z{}^2$. However, elements of the form $\M{\ell}{0}$ and $\M{0}{m}$ 
build two independent subgroups of \Sp{4} because
\begin{subequations}
\begin{eqnarray}
\M{\ell_1}{0}\,\M{\ell_2}{0} & = & \M{\ell_1+\ell_2}{0}\;,\\
\M{0}{m_1}\,\M{0}{m_2}       & = & \M{0}{m_1+m_2}\;.
\end{eqnarray}
\end{subequations}

Next, we consider the action of $\gamma\in\SL{2,\Z{}}_{U}$ on $M_{\Delta}$. It is given by
\begin{equation}
M_{(\Id_2, \gamma)}\,M(\Delta) ~=~ M(\gamma\,\Delta)\,M_{(\Id_2,\gamma)}
\end{equation}
such that for $\Delta = (\ell,m)^\mathrm{T}$ we obtain
\begin{subequations}\label{eq:SandTonWL}
\begin{eqnarray}
M_{(\Id_2, \mathrm{S})}\,M\!\left(\!\begin{smallmatrix}\ell\\m\end{smallmatrix}\!\right) & = & M\!\left(\!\begin{smallmatrix}m\\-\ell\end{smallmatrix}\!\right)\,M_{(\Id_2, \mathrm{S})}\;,\\
M_{(\Id_2, \mathrm{T})}\,M\!\left(\!\begin{smallmatrix}\ell\\m\end{smallmatrix}\!\right) & = & M\!\left(\!\begin{smallmatrix}\ell+m\\m\end{smallmatrix}\!\right)\,M_{(\Id_2, \mathrm{T})}\;.
\end{eqnarray}
\end{subequations}
From the point of view of Wilson lines on a two-torus, these equations are not unexpected: $\ell$ 
corresponds to the Wilson line $A_2$ in the $e_2$ direction, while $m$ corresponds to the Wilson 
line $A_1$ in the $e_1$ direction. Furthermore, under modular $\mathrm{S}$ and $\mathrm{T}$ 
transformations from $\SL{2,\Z{}}_{U}$, the lattice vectors $e_1$ and $e_2$ get mapped as
\begin{equation}\label{eq:SandTonei}
e_1 \stackrel{\mathrm{S}}{\rightarrow} -e_2\;,\;\; e_2 \stackrel{\mathrm{S}}{\rightarrow} e_1\;,\;\;\mathrm{and}\;\;e_1 \stackrel{\mathrm{T}}{\rightarrow} e_1\;,\;\; e_2 \stackrel{\mathrm{T}}{\rightarrow} e_1+e_2\;,
\end{equation}
see e.g.\ ref.~\cite{Nilles:2020gvu}. Hence, eqs.~\eqref{eq:SandTonWL} resembles eq.~\eqref{eq:SandTonei} 
on the level of the associated Wilson lines.
In addition, we have checked the following relations, both in \Sp{4} and in 
$\mathrm{O}_{\hat\eta}(2,2+16,\Z{})$:
\begin{subequations}
\begin{eqnarray}
M_{(\mathrm{T}, \Id_2)}\,\M{\ell}{m} \left(M_{(\mathrm{T}, \Id_2)}\right)^{-1}
                                             & = & \M{\ell}{m}\;,\\ 
M_\times\,M_{(\gamma_1, \gamma_2)}\,M_\times & = & M_{(\gamma_2, \gamma_1)}\;,\\
M_\times\,\M{\ell}{0}\, M_\times             & = & \M{\ell}{0}\;,\\
M_\times\,\M{0}{m}\, M_\times                & = & \left(M_{(\mathrm{S}, \mathrm{S})}\right)^{-1} \M{0}{-m}\, M_{(\mathrm{S}, \mathrm{S})}\;.
\end{eqnarray}
\end{subequations}
Finally, we learn from the relation
\begin{equation}
\M{\ell}{m} ~=~ \M{\ell}{0}\,\left(M_{(\Id_2, \mathrm{S})}\right)^{-1}\M{m}{0}\,M_{(\Id_2, \mathrm{S})}\,
                M_\times\, \left(M_{(\Id_2, \mathrm{T})}\right)^{-\ell m}\,M_\times
\end{equation}
that \Sp{4} can be generated by $M_{(\Id_2, \mathrm{S})}$, $M_{(\Id_2, \mathrm{T})}$, $M_\times$ 
and $\M{1}{0}$ (and its inverse $\M{-1}{0}$).



\providecommand{\bysame}{\leavevmode\hbox to3em{\hrulefill}\thinspace}

\end{document}